\documentclass[twocolumn,showpacs,preprintnumbers,prc]{revtex4}
\usepackage{graphicx}
\newcommand{\be}{\begin{equation}}
\newcommand{\ee}{\end{equation}}
\newcommand{\bea}{\begin{eqnarray}}
\newcommand{\eea}{\end{eqnarray}}
\newcommand{\nn}{\nonumber}

\newcommand{\nl}{\newline}

\begin{document}
\title{\bf{Examination of the Gunion-Bertsch formula for soft gluon radiation}}
\author{Trambak Bhattacharyya\footnote{trambakb@vecc.gov.in}, 
Surasree Mazumder, Santosh K Das and Jan-e Alam}
\medskip
\affiliation{Theoretical Physics Division, 
Variable Energy Cyclotron Centre, 1/AF, Bidhannagar, Kolkata-700064}
\date{\today}
\begin{abstract}
The spectrum of emitted gluons from the process $\mathrm{g+g\rightarrow g+g+g}$ 
has been evaluated by relaxing some of the approximations used in earlier works.
The difference in the results from earlier calculations have  been pointed out.
The formula obtained in the present work has been applied to estimate physical 
quantities like equilibration rate of gluons and the energy loss of fast gluon 
in the gluonic plasma. 
\end{abstract}

\maketitle

%%%%%%%%%%%%%%%%%%%%%%%%%%%%%%%%%%%%%%%%%%%%%%%%%%%%%%%%%%%%%%%%%%%%%%%%%%%%%%%%%%%%%%%%%%%%%%%%%%%%%%%%%%%%%%%%%%%%%%%%%%%%
%\section{Introduction}
The radiation of soft gluons from the generic partonic ($\cal{P}$) 
processes $\cal{P}$$_1$+$\cal{P}$$_2$ $\rightarrow$$\cal{P}$$_3$+
$\cal{P}$$_4$+gluon play a crucial role in the study of 
quark gluon plasma (QGP)
expected to be formed in heavy ion collisions (HIC)
at ultra-relativistic energies.  
The number nonconserving process, $\mathrm{g+g\rightarrow g+g+g}$ has drawn 
particular attention in view of its importance for the 
(i) chemical equilibration in the deconfined phase of quarks and gluons 
~\cite{biro,xiong,zxu}, (ii) energy loss of fast gluons propagating 
through gluonic plasma~\cite{wgp,fxg}, (iii) evaluation of transport 
coefficients 
of the gluonic plasma~\cite{XG,CDOW,amy,DAvis}, etc. 
The Gunion-Bertsch (GB)~\cite{GB} formula for the spectrum of the radiated 
gluons from the processes $\mathrm{g+g\rightarrow g+g+g}$
has been widely used in many of these calculations. 
Recently, attempts have been made to revisit/generalize the GB formula
~\cite{DA,GR}. 
The main purpose of the present work is to find the 
corrections to the GB formula by relaxing some of the 
approximations previously adopted.  Then, we inspect the effects of the 
correction terms  relative to the GB formula on some physical quantities like 
the equilibration time  and the energy loss of fast 
gluons in a gluonic plasma.
The results obtained here can be extended 
to  other partonic processes  such as
$q + q\rightarrow q + q + \mathrm{g}$, 
$q + \mathrm{g}\rightarrow q + \mathrm{g} + \mathrm{g}$, etc.,  where $q$ 
stands for quark.

We consider the process
$\mathrm{g}(k_1) + \mathrm{g}(k_2)\rightarrow \mathrm{g}(k_3) + \mathrm{g}(k_4) + \mathrm{g}(k_5)$.
The square of the invariant amplitude for this reaction
can be written elegantly  as~\cite{berendes}:
%%%%%%%%%%%%%%%%%%%%%%
%\begin{figure}[h]
%\begin{center}
%\includegraphics[scale=0.50]{gg2gggpic.eps}
%\end{center}
%\caption{Feynman diagram of $gg\rightarrow ggg$ process.}
%\label {fig.1}
%\end{figure}
\bea
|M_{\mathrm{gg\rightarrow ggg}}|^2&=&\frac{1}{2}g^6 \frac{N_c ^3}{N_c ^2-1}\frac{\mathcal{N}}
{\mathcal{D}} \nn\\
&\times&[(12345)+(12354)+(12435)\nn\\
&+&(12453)+(12534)+(12543)\nn\\
&+&(13245)+(13254)+(13425)\nn\\
&+&(13524)+(14235)+(14325)],
\eea
where
\bea
\mathcal{N}&=&(k_1.k_2)^4+(k_1.k_3)^4+(k_1.k_4)^4\nn\\
           &+&(k_1.k_5)^4+(k_2.k_3)^4+(k_2.k_4)^4+(k_2.k_5)^4\nn\\
           &+&(k_3.k_4)^4+(k_3.k_5)^4+(k_4.k_5)^4;
\eea
\bea
\mathcal{D}&=&(k_1.k_2)(k_1.k_3)(k_1.k_4)(k_1.k_5)(k_2.k_3)\nonumber\\
&\times&(k_2.k_4)(k_2.k_5)(k_3.k_4)(k_3.k_5)(k_4.k_5),
\eea
and
\be
(ijklm)=(k_i.k_j)(k_j.k_k)(k_k.k_l)(k_l.k_m)(k_m.k_i).
\ee
$N_c(=3)$ is the number of colors, $g=\sqrt{4\pi\alpha_s}$ is the color charge, and
$\alpha_s$ is the strong coupling.

The quantity, $|M_{\mathrm{gg\rightarrow ggg}}|^2$
after simplification can be written as (see Appendix):
\bea
{|M|^2}_{\mathrm{gg\rightarrow ggg}}&=& 12g^2 |{M _{\mathrm{gg\rightarrow gg}}}|^2 \frac{1}{k_{\bot}^2} \nonumber\\
&\times&[(1+\frac{t}{2s}+\frac{5t^2}{2s^2}-\frac{t^3}{s^3})\nonumber\\
&-&(\frac{3}{2\sqrt{s}}+\frac{4t}{s\sqrt{s}}-\frac{3t^2}{2s^2\sqrt{s}})k_{\bot} \nonumber\\
&+&(\frac{5}{2s}+\frac{t}{2s^2}+\frac{5t^2}{s^3})k_{\bot}^2],
\label{matele}
\eea
where
$|M_{\mathrm{gg\rightarrow gg}}|^2=(9/2)g^4s^2/t^2$,
$s=(k_1+k_2)^2, t=(k_1-k_3)^2$, $u=(k_1-k_4)^2$,
$k_{\perp}$ is the transverse momentum of the radiated gluon.
%The derivation of Eq.~\ref {matele} is outlined in the appendix.
The  $\mathcal{O}(k_\perp^{-1})$ and 
 $\mathcal{O}(k_\perp^0)$  terms appearing in Eq.~\ref{matele} 
were absent in Ref.~\cite{GR}. Henceforth these two terms
will be called the correction terms.  We will demonstrate that
the contributions from these terms  are non-negligible and 
will have  crucial importance for the
phenomenology of heavy ion collisions at ultra-relativistic energies.

%Next we apply our results to evaluate the energy loss of the
%fast gluons propagating through QGP.
While the details for the derivation of the Eq.~\ref{matele} 
is given in the Appendix,  we would  like to  
check the effects of the correction terms in Eq.~\ref{matele} 
to physical quantities like  equilibration time 
of gluons and energy loss of fast gluons propagating through a
gluonic fluid. 

\begin{figure}[h]                                                            %
\begin{center}                                                               %
\includegraphics[scale=0.40]{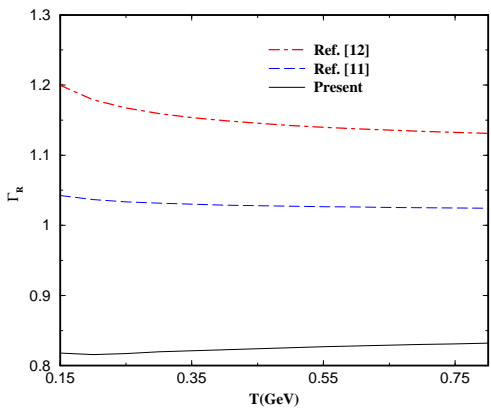}                                 %
\end{center}                                                                 %
\caption{(Color online) 
Temperature variation of the ratio of the equilibration rate  
(inverse of the time scale) 
obtained in the present work (solid line), Ref.~\cite{DA} (dashed line),
and  ~\cite{GR} (dot-dashed) normalized 
by the GB value 
for the process $\mathrm{gg}\rightarrow \mathrm{ggg}$. 
}
\label {fig1}                                                               %
\end{figure}

Let us first discuss the role of the correction terms 
in the equilibrium time scale 
of gluons. The processes $\mathrm{g}+\mathrm{g}\leftrightarrow 
\mathrm{g}+\mathrm{g}$, $\mathrm{g}+\mathrm{g}\leftrightarrow 
\mathrm{g}+\mathrm{g}+\mathrm{g}$,
$\mathrm{g}+\mathrm{g}\leftrightarrow 
\mathrm{g}+\mathrm{g}+\mathrm{g}+\mathrm{g}$ etc are
responsible for maintaining equilibration in the system. 
In the present work we would like to estimate
the contribution from the correction terms of
the process $\mathrm{g}+\mathrm{g}\leftrightarrow 
\mathrm{g}+\mathrm{g}+\mathrm{g}$ as mentioned above. 
We evaluate the equilibration rate for this process 
(with $s=18T^2$ as in ~\cite{biro}) 
with and without the correction terms.  The 
equilibration rates obtained from the spectra of Refs.~\cite{DA},\,~\cite{GR}
and present work normalized by the GB spectra 
($\Gamma_R$)  are displayed in Fig.~\ref{fig1}.
We observe that the equilibration rate obtained with 
the correction terms is  smaller  
compared to the scenario when the
corrections are neglected for the entire range
of temperature under consideration. 
This indicates that the contribution from
the correction terms will enhance the equilibration in the
gluonic system.

Before going into estimating the energy loss
of fast partons moving through a gluonic fluid,
some clarification are in order here. 
The energy loss of fast 
partons in QGP has been evaluated rigorously (see~\cite{rchwavol,annrev} 
for review).  In the present work the aim is not to
achieve the same level of rigor but to check the effects of
the correction terms discussed before on the energy loss
of fast partons. 

The energy loss of high-energy partons propagating through QGP
is a field of high contemporary interest. 
Experimentally the energy dissipation has
been measured  through the suppression of the transverse  momentum ($p_T$)
distribution of hadrons produced in Au+Au collision 
relative to the binary scaled p+p interaction  at
the same center of mass energy. The nature of the suppression may be used as a
tool for diagnosis of QGP formation in nuclear collisions at Relativistic
Heavy Ion Collider (RHIC) and Large Hadron Collider (LHC).
The two most common mechanisms for
the energy loss are elastic and inelastic or radiative processes. Among 
various in-elastic processes involving quarks and gluons, the process:
$\mathrm{g} + \mathrm{g}\rightarrow \mathrm{g} + \mathrm{g} + \mathrm{g}$ plays a major role.

\begin{figure}[h]                                                            %
\begin{center}                                                               %
\includegraphics[scale=0.40]{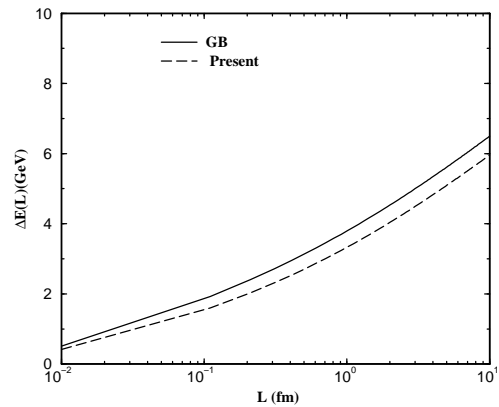}                                 %
\end{center}                                                                 %
\caption{Energy loss of a 15 GeV gluon in vacuum 
as a function of path length. 
}
\label {fig1a}                                                               %
\end{figure}

In the hard processes the gluons produced with virtuality $\sim Q^2$ is
highly off-shell because during the collision process the 
color field of the gluon is stripped off. Therefore, the highly virtual gluon
will develop a dead cone - which will results in radiative suppression. Following
the procedure outlined in Ref~\cite{bzk} we evaluate the energy loss of gluon,
$\Delta E(L)$  as a function of length, $L$. The results 
obtained with GB and present gluon spectra are displayed in Fig.~\ref{fig1a}.
We observe that the difference in $\Delta E(L)$ between the GB spectra and the
spectra derived in the present work (with corrections described above) is
small at lower $L$ but grows up  with increase in $L$.  

The spectrum of the radiated gluon in medium,
derived by using the ratio of the amplitude square of the radiative process
$\mathrm{gg} \rightarrow \mathrm{ggg}$ to that of the
elastic process, $\mathrm{gg} \rightarrow \mathrm{gg}$ is given by,
\bea
\frac{dn_g}{d^2 k_{\perp}d\eta}&=&\left[\frac{dn_g}{d^2 k_{\perp}d\eta}\right]_ {GB}
[(1+\frac{t}{2s}+\frac{5t^2}{2s^2}-\frac{t^3}{s^3})\nn\\
&-&(\frac{3}{2\sqrt{s}}+\frac{4t}{s\sqrt{s}}-\frac{3t^2}{2s^2\sqrt{s}})k_{\bot} \nn\\
&+&(\frac{5}{2s}+\frac{t}{2s^2}+\frac{5t^2}{s^3})k_{\bot}^2)],
\label{spectrum}
\eea
where $\eta$ is the rapidity of the radiated gluon, the subscript GB has been used to indicate the
gluon spectrum obtained using the approximation considered in~\cite{GB} (see
also~\cite{wong})
which is given by,
\be
\left[\frac{dn_g}{d^2 k_{\perp}d\eta}\right]_ {GB}=\frac{C_A \alpha_s}{\pi^2}\frac{q_{\perp}^2}{k_{\perp}^2[({\bf k}_{\perp}-{\bf q}_{\perp})^2
+m_D^2]},
\label{gbspectrum}
\ee
where $m_ D=\sqrt{2 \pi\alpha_s(T)(C_A+\frac{N_F}{2})T/3}$,
is the thermal mass of the gluon~\cite{Bellac}, $N_F$ is the number of flavors contributing
in the gluon self-energy loop,
$C_A=3$ is the Casimir invariant for the SU(3) adjoint representation,
$\alpha_s$ is the temperature-dependent strong coupling~\cite{KACZ}
and $q_{\perp}$ is the transverse
momentum transfer.
The thermal mass in the denominator of Eq.~\ref{gbspectrum}
has been introduced to shield the infrared divergence arising from
the massless intermediary gluon exchange.

%The momentum spectrum of the radiated gluon  can be used to estimate
%the energy loss of fast partons propagating in QGP.
%The momentum distribution of the radiated gluons from the above process,
%considered by Gunion and Bertsch (GB)~\cite{GB} long ago,
%has drawn some attention recently~\cite{DA,GR}.

The energy loss of a gluon passing through a gluonic medium can now
be calculated using the gluon spectrum of Eq.~\ref{spectrum}.~The 
energy loss per collision can now be estimated as:
\bea
\epsilon&=&\int d^2 k_{\perp} d\eta \frac{dn_g}{d^2 k_{\perp}d\eta} k_0 \theta(\Lambda^{-1} -\tau_F)\nn\\
&\times& \theta(E-k_{\perp}cosh\eta),
\label{eloss}
\eea
where $k_0=k_{\perp} cosh\eta$ is the energy of the radiated gluon
and $\tau_F$ is the formation time of the gluon.
The first $\theta$-function in Eq.~\ref{eloss}, involving $\Lambda^{-1}$ or interaction time,
is introduced for the Landau-Pomeranchuk-Migdal (LPM) effect. 
%The effect is actually due
%to a characteristic destructive interference phenomenon caused by finite formation time of the
%radiated gluon with 4-momentum $k_5=(k_0,k_{\bot},k_z)$ defined by
%$\tau_F\sim {1}/{\Delta E}$,
%where $\Delta E$ is the energy lost by the particle in a single collision.
%In effect,  $\tau_F $ is the minimum time needed to resolve the transverse wave-packet of the quanta with
%$\Delta x_{\bot} \sim {1}/{k_{\bot}}$ from its high energy parent($E>>k_0$). 
%%It is the time within which another interaction, if occurs, results in the suppression of the emission of the gluon.
%Destructive interference among the radiation amplitudes associated with multiple scattering is expected when
%the formation time is larger than the mean free path, $\lambda$. When $\tau_F>>\lambda$ the scattering centers cannot
%resolve the emitted quanta and the incoherent contribution of each scattering breaks down. This effect is called
%Landau-Pomeranchuk-Migdal (LPM) suppression. 
The LPM effect imposes restriction
on the phase-space of the radiated gluon,
it  must have $\tau_F (=cosh\eta/k_{\perp})$ 
less than the mean free time, $\Lambda^{-1}$. 
The gluon can be emitted for time scales  larger than $\tau_F$~\cite{wgp}.
The second $\theta$-function sets the upper limit for the energy of the
radiated gluon.

\begin{figure}[h]                                                            %
\begin{center}                                                               %
\includegraphics[scale=0.40]{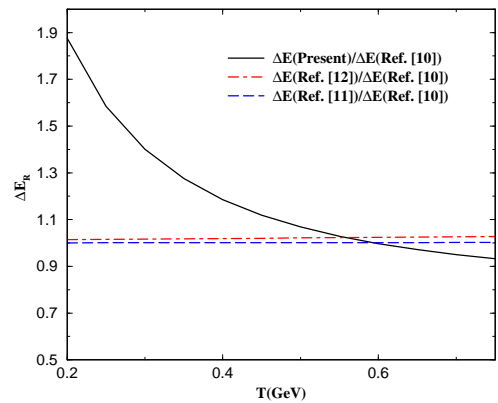}                                 %
\end{center}                                                                 %
\caption{(Color online) Temperature variation of $\Delta E_R$ of a 15 GeV gluon 
moving through a gluonic heat bath of dimension 4 fm. Solid (dashed) line indicates result for
the gluon spectrum obtained in the present work (~\cite{DA}). 
The dot-dashed line stands for the results 
for the gluon spectrum of ~\cite{GR}. The $\Delta E$  normalized by 
the corresponding value obtained from GB approximation and 
the resulting quantity is called $\Delta E_R$. 
}                                                          %
\label {fig2}                                                               %
\end{figure}

To proceed further, we replace
$q_{\perp}^2$ by its average value evaluated as follows:

\be
\langle q_{\perp}^2 \rangle=\frac{1}{\sigma_{el}} \int_ {m_D ^2} ^\frac{s}{4} dq_{\perp}^2 \frac{d\sigma_{el}}
{dq_{\perp}^2}q_{\perp}^2,
\ee
where

\be
\sigma_{el}=\int_ {m_D ^2} ^\frac{s}{4} dq_{\perp}^2 \frac{d\sigma_{el}}{dq_{\perp}^2}.
\ee
For dominant small-angle scattering ($t\rightarrow 0$),
\be
\frac{d\sigma_{el}}{dq_{\perp}^2}=C_i \frac{2\pi \alpha_s ^2}{q_{\perp}^4}.
\ee
$C_i$ is $9/4$, 1 and $4/9$ for $\mathrm{gg}$, $\mathrm{qg}$, and qq scattering.
$\langle q_{\perp}^2 \rangle$ is then obtained as,
\be
\langle q_{\perp}^2 \rangle=\frac{s m_D ^2}{s-4m_D ^2}\ln (\frac{s}{4m_D ^2}).
\label{qperp}
\ee
For  $\sqrt{s}\rightarrow\infty$, i.e. in the high-energy limit
one can make the replacement $t\sim -q_{\perp} ^2$ ~\cite{PRE}.
In contrast to previous works~\cite{DA,GR} the value of $s$ is taken as
$s\sim 6ET$ ~\cite{wgp} allowing the possibility for the incident gluon to remain 
out of thermal equilibrium. 
%we put $s\sim6ET$ in Eq.~\ref{qperp}, allowing the
%possibility for the incident gluon to remain out of thermal 
%equilibrium ($E\neq 3T$).
With all the above ingredients we are now ready to evaluate
energy loss ($dE/dx$) of a fast gluon in a gluonic medium as follows:
\be
-\frac{dE}{dx}=\epsilon\cdot\Lambda.
\label{dedx}
\ee
The interaction rate, $\Lambda$ has been evaluated by
using the procedure similar to~\cite{thoma}.
We need to implement now the  radiative suppression 
of gluons due to its possible off-shellness (Fig.~\ref{fig1a}). 
The energy loss  ($\Delta E$) of off-shell gluons of energy 15 GeV moving 
through a gluonic fluid of dimension 4 fm for the gluon spectrum derived in Refs.~\cite{DA},~\cite{GR} and
in the present work [obtained from Eqs.~\ref{spectrum}, ~\ref{eloss} and ~\ref{dedx}]
normalized to that resulting  from  GB approximations $(\Delta E_R)$ are displayed in Fig.~\ref{fig2}.
We observe that with the correction terms the value of $\Delta E$ is
enhanced by about $40\%$ and $20\%$ for $T$ 300 MeV and 400 MeV, respectively,
compared to the $\Delta E$ obtained from the spectra of  Refs.~\cite{DA} and ~\cite{GR}.
Such differences may have important consequences on the
heavy-ion phenomenology at RHIC and LHC collision energies.

%\section{RESULTS AND DISCUSSION}
In summary we have evaluated the spectrum of the emitted gluon
from the processes $\mathrm{g+g\rightarrow g+g+g}$ by relaxing some
of the approximations used in recent calculations. 
The results derived in the present calculation has been
applied to evaluate some physical quantities, {\it e.g.} 
energy loss of energetic partons in HIC and time scale
for  equilibration in gluonic plasma. Results
obtained in the present work have been
compared with the earlier calculations~\cite{DA,GR}.
We find that the contributions from the 
previously neglected terms $\mathcal{O}(k_\perp^{-1})$ and
$\mathcal{O}(k_\perp^{0})$ in the matrix element 
play crucial role in heavy-ion phenomenology.
The gluon spectrum obtained in Eq.~\ref{spectrum} can be used
for other $\mathrm{2\,\rightarrow 2 + g}$ partonic processes also.

%*****************************************************************************
%*****************************************************************************
%*****************************************************************************
%*****************************************************************************
%\begin{figure}[h]                                                            %
%\begin{center}                                                               %
%\includegraphics[scale=0.50]{fig3.eps}                                    %
%\end{center}                                                                 %
%\caption{(color online)
%Variation of $dE/dx$  (normalized by the $dE/dx$
%for GB approximation) with $T$. Solid line indicates the result
%for the present work. Dashed (dotted) line depicts the energy loss
%obtained from the gluon spectrum of Ref.~\cite{GR} (~\cite{DA}).
%}
%\label {fig3}                                                               %
%\end{figure}                                                                 %
%*****************************************************************************

%%%%%%%%%%%%%%%%%%%%%%%%%%%%%%%%%%%%%%%%%%%%%%%%%%%%%%%%%%%%%%%%%%%%%%%%%%%%%%%%%%%%%%%%%%%%%%%%%%%%%%%%%%%%%%%%%%%%%%%%
\section*{Appendix}
%%%%%%%%%%%%%%%%%%%%%%%%%%%%%%%%%%%%%%%%%%%%%%%%%%%%%%%%%%%%%%%%%%%%%%%%%%%%%%%%%%%%%%%%%%%%%%%%%%%%%%%%%%%
In this appendix we derive Eq.~\ref{matele} for the
square of the invariant amplitude for the
radiative process, $\mathrm{gg\rightarrow ggg}$ upto orders
$\mathcal{O}({\mathrm{k}_{\perp}}^0)$ and
$\mathcal{O}(\frac{\mathrm{t}^3}{\mathrm{s}^3})$.
Consider the reaction
\be
\mathrm{g}(k_1) + \mathrm{g}(k_2)\rightarrow \mathrm{g}(k_3) + \mathrm{g}(k_4) + \mathrm{g}(k_5),
\ee
where $k_5$ is the four-momentum of the radiated gluon.
The Mandelstam variables for the  above process are defined as
\bea
s&=&(k_1+k_2)^2,\,\,\,\, t=(k_1-k_3)^2\nonumber\\
u&=&(k_1-k_4)^2,\,\,\,\, s^{\prime}=(k_3+k_4)^2\nonumber\\
t^{\prime}&=&(k_2-k_4)^2,\,\,\,\, u^{\prime}=(k_2-k_3)^2.
\eea
Because gluons massless we can write
\bea
k_1.k_2&=&\frac{s}{2},\,\,\,\, k_1.k_3=-\frac{t}{2}\nonumber\\
k_1.k_4&=&-\frac{u}{2},\,\,\,\, k_3.k_4=\frac{s^{\prime}}{2}\nonumber\\
k_2.k_4&=&-\frac{t^{\prime}}{2},\,\,\,\, k_2.k_3=-\frac{u^{\prime}}{2}.
\eea
We also have the relations
\bea
k_1.k_5&=&\frac{s+t+u}{2},\,\,\,\,k_2.k_5=\frac{s+t^{\prime}+u^{\prime}}{2}\nonumber\\
k_3.k_5&=&\frac{s+t^{\prime}+u}{2},\,\,\,\, k_4.k_5=\frac{s+t+u^{\prime}}{2}.
\eea
For soft gluon emission,
\be
s+t+u+s^{\prime}+t^{\prime}+u^{\prime}=0.
\ee
The matrix element square of the radiative process
$\mathrm{gg\rightarrow ggg}$ is given by ~\cite{berendes}

\bea
|M_{\mathrm{gg\rightarrow ggg}}|^2&=&\frac{1}{2}g^6 \frac{N_c ^3}{N_c ^2-1}\frac{\mathcal{N}}{\mathcal{D}} \nonumber\\
 &\times&[(12345)+(12354)+(12435)\nonumber\\
&+&(12453)+(12534)+(12543)\nonumber\\
&+&(13245)+(13254)+(13425)\nn\\
&+&(13524)+(14235)+(14325)],
\label{eq19}
\eea
where $N_c$ is the number of colors, $g=\sqrt{4\pi\alpha_s}$ is the strong coupling,
\bea
\mathcal{N}&=&(k_1.k_2)^4+(k_1.k_3)^4+(k_1.k_4)^4\nn\\
&+&(k_1.k_5)^4+(k_2.k_3)^4+(k_2.k_4)^4+(k_2.k_5)^4\nn\\
&+&(k_3.k_4)^4+(k_3.k_5)^4+(k_4.k_5)^4,
\eea
\bea
\mathcal{D}&=&(k_1.k_2)(k_1.k_3)(k_1.k_4)(k_1.k_5)(k_2.k_3)\nonumber\\
&\times&(k_2.k_4)(k_2.k_5)(k_3.k_4)(k_3.k_5)(k_4.k_5),
\eea
and
\be
(ijklm)=(k_i.k_j)(k_j.k_k)(k_k.k_l)(k_l.k_m)(k_m.k_i).
\ee
Simplifying Eq.~\ref{eq19} we get,

\bea
|M_{\mathrm{gg\rightarrow ggg}}|^2&=& 16g^6 \frac{N_c ^3}{N_c ^2-1}\mathcal{N}\nonumber\\
&\times& [\frac{1} {s^{\prime}(s+u+t)(s+u^{\prime}+t^{\prime})}[\frac{1} {tt^{\prime}}+\frac{1}{uu^{\prime}}]\nonumber\\
&+&\frac{1}{s(s+u+t^{})(s+u^{\prime}+t)}[\frac{1}{tt^{\prime}}+\frac{1}{uu^{\prime}}]\nonumber\\
&-&\frac{1}{t^{\prime}(s+u+t)(s+u+t^{\prime})}[\frac{1}{uu^{\prime}}+\frac{1}{ss^{\prime}}]\nonumber\\
&-&\frac{1}{u^{\prime}(s+u+t)(s+u^{\prime}+t)}[\frac{1}{tt^{\prime}}+\frac{1}{ss^{\prime}}]\nonumber\\
&-&\frac{1}{u(s+u^{\prime}+t^{\prime})(s+u+t^{\prime})}[\frac{1}{tt^{\prime}}+\frac{1}{ss^{\prime}}]\nonumber\\
&-&\frac{1}{t(s+u^{\prime}+t^{\prime})(s+u^{\prime}+t)}[\frac{1}{uu^{\prime}}\nn\\
&+&\frac{1}{ss^{\prime}}];
\label {mateleberendes}
\eea
and $\mathcal{N}$ can now be written as

\bea
\mathcal{N}&=&\frac{1}{16} [s^4+t^4+u^4+{s^{\prime}}^4+{t^{\prime}}^4+{u^{\prime}}^4\nonumber\\
&+&(s+t+u)^4+(s+t^{\prime}+u^{\prime})^4+(s+t^{\prime}+u)^4\nn\\
&+&(s+t+u^{\prime})^4].
\label {NBerendes}
\eea
\nl
\nl
For a soft gluon emission ($k_5\rightarrow 0$) $s\rightarrow s^{\prime}$, $t\rightarrow t^{\prime}$,
$u\rightarrow u^{\prime}$. We can express the transverse momentum of the emitted
gluon as

\bea
k_{\bot} ^2&=& 4(k_1.k_5)(k_2.k_5)/s \nonumber\\
           &=& (s+t+u)(s+t^{\prime}+u^{\prime})\nonumber\\
           &=& (s+t+u)^2/s\nonumber\\.
\label {kperpsq}
\eea
Using Eqs.~\ref{mateleberendes},~\ref{NBerendes} and \ref{kperpsq}, the
square of the matrix element can be written as
%\bea
%{|M|^2}_{\mathrm{gg\rightarrow ggg}}&=&\frac{27}{2} g^6 [s^4+t^4+u^4+(s+t+u)^4]\nn\\
%&\times&\frac{1}{(s+t+u)^2}[\frac{1}{s}(\frac{1}{t^2}+\frac{1}{u^2})\nonumber\\
%&-&\frac{1}{t}(\frac{1}{s^2}+\frac{1}{u^2})\nonumber\\
%&-&\frac{1}{u}(\frac{1}{t^2}+\frac{1}{s^2})]
%\label {mateleberendes1}
%\eea
%Again, putting equation ~\ref{kperpsq} in ~\ref{mateleberendes1} we get,
\bea
{|M|^2}_{\mathrm{gg\rightarrow ggg}}&=&\frac{27}{2} g^6 (s^4+t^4+u^4+2s^2k_{\bot} ^4)\frac{1}{sk_{\bot} ^2} \nonumber\\
&\times&[\frac{1}{s}(\frac{1}{t^2}+\frac{1}{u^2})\nonumber\\
&-&\frac{1}{t}(\frac{1}{s^2}+\frac{1}{u^2})\nonumber\\
&-&\frac{1}{u}(\frac{1}{t^2}+\frac{1}{s^2})]\nonumber\\
%%%%%%%%%%%%%%%%%%%%%%%%%%%%%%%%%%%%%%%%%%%%%%%%%%%%%%%%%%%%%%%%%%%%%%%%%%%%%%%%%%%%%%%%%%%%%%%%%%%%%%%%%%%%%%%%%%%%%%%%%
&=&g^2 (\frac{27}{2}g^4s^4) (1+\frac{t^4}{s^4}+\frac{u^4}{s^4}+2\frac{k_{\bot} ^4}{s^2}) \nonumber\\
&\times&\frac{1}{s^2 k_{\bot} ^2 t^2}[1+\frac{t^2}{u^2}-\frac{t}{s}-\frac{st}{u^2}-\frac{s}{u}-\frac{t^2}{us}] \nonumber\\
%%%%%%%%%%%%%%%%%%%%%%%%%%%%%%%%%%%%%%%%%%%%%%%%%%%%%%%%%%%%%%%%%%%%%%%%%%%%%%%%%%%%%%%%%%%%%%%%%%%%%%%%%%%%%%%%%%%%%%%%%
&=&g^2(\frac{9}{2}g^4 \frac{s^2}{t^2})(3+3\frac{t^4}{s^4}+3\frac{u^4}{s^4}+6\frac{k_{\bot} ^4}{s^2})\frac{1}{k_{\bot}^2}\nonumber\\
&\times&[1+\frac{t^2}{u^2}-\frac{t}{s}-\frac{st}{u^2}-\frac{s}{u}-\frac{t^2}{us}]\nonumber\\
%%%%%%%%%%%%%%%%%%%%%%%%%%%%%%%%%%%%%%%%%%%%%%%%%%%%%%%%%%%%%%%%%%%%%%%%%%%%%%%%%%%%%%%%%%%%%%%%%%%%%%%%%%%%%%%%%%%%%%%%%
&=&g^2 (\frac{9}{2}g^4 \frac{s^2}{t^2}) (3(1+\frac{u^4}{s^4})+3\frac{t^4}{s^4}+6\frac{k_{\bot} ^4}{s^2}) \nonumber\\
&\times&\frac{1}{k_{\bot} ^2}[1-\frac{s}{u}- (1+\frac{s^2}{u^2})\frac{t}{s}+ (\frac{s^2}{u^2}-\frac{s}{u})\frac{t^2}{s^2}]\nonumber\\
%%%%%%%%%%%%%%%%%%%%%%%%%%%%%%%%%%%%%%%%%%%%%%%%%%%%%%%%%%%%%%%%%%%%%%%%%%%%%%%%%%%%%%%%%%%%%%%%%%%%%%%%%%%%%%%%%%%%%%%%%
&=&g^2 |{M _{\mathrm{gg\rightarrow gg}}}|^2 (3(1+\frac{u^4}{s^4})+3\frac{t^4}{s^4}+6\frac{k_{\bot} ^4}{s^2}) \nonumber\\
&\times&\frac{1}{k_{\bot} ^2}[(1-\frac{s}{u})- (1+\frac{s^2}{u^2})\frac{t}{s}\nn\\
&+&(\frac{s^2}{u^2}-\frac{s}{u})\frac{t^2}{s^2}],
\label{eq27}
\eea
where the subscript GB stands for the approximation used by Gunion and Bertsch~\cite{GB}.
For the elastic process,
\be
|{M _{\mathrm{gg\rightarrow gg}}}|^2=\frac{9}{2}g^4\frac{s^2}{t^2}.
\ee
On simplifying Eq.~\ref{eq27} we obtain,

\bea
{|M|^2}_{\mathrm{gg\rightarrow ggg}}&=& g^2 |{M _{\mathrm{gg\rightarrow gg}}}|^2\nonumber\\
&\times&\frac{1}{k_{\bot}^2} [(3-3\frac{s}{u}+3\frac{u^4}{s^4}-3\frac{u^3}{s^3})- (3\frac{t}{s}+3\frac{st}{u^2}\nonumber\\
&+&3\frac{u^4 t}{s^5}+3\frac{u^2 t}{s^3})\nonumber\\
&+& (3\frac{t^2}{u^2}-3\frac{t^2}{us}+3\frac{u^2t^2}{s^4}-3\frac{u^3t^2}{s^5})\nn\\ &+&(3\frac{t^4}{s^4}-3\frac{t^4}{us^3})\nonumber\\
&-& (3\frac{t^5}{s^5}+3\frac{t^5}{u^2s^3})+ (3\frac{t^6}{u^2s^4}-3\frac{t^6}{us^5})\nonumber\\
&+& (6\frac{k_{\bot} ^4}{s^2}-6\frac{k_{\bot} ^4}{us})- (6\frac{k_{\bot} ^4t}{s^3}+6\frac{k_{\bot} ^4t}{u^2s})\nonumber\\
&+& (6\frac{k_{\bot}^4 t^2}{u^2s^2}-6\frac{k_{\bot}^4 t^2}{us^3})].
\label {matelsq}
\eea
In the proposed kinematic limit we set terms which are linear in $k_{\bot}$ to zero and keep terms
$\mathcal{O}({k_{\perp}^0}), \mathcal{O}({k_{\perp}^{-1}})$, and $\mathcal{O}({k_{\perp}^{-2}})$
in ${|M|^2}_{\mathrm{gg\rightarrow ggg}}$. We also neglect
terms $\mathcal{O}(\frac{t^4}{s^4})$ and higher order in the matrix element.
To proceed further one needs to express the Mandelstam variable $u$ in terms of
$s$, $t$, and $k_{\bot}$ by using the following relation:
\bea
{k_{\bot}}^2&=&\frac{(s+t+u)^2}{s} \nonumber\\
\Rightarrow u&=&\sqrt{s} k_{\bot}-s-t \nonumber\\
\Rightarrow \frac{1}{u}&=&\frac{1}{(\sqrt{s} k_{\bot}-s-t)} \nonumber\\
\Rightarrow \frac{1}{u}&=&-\frac{1}{s} [1-(\frac{ k_{\bot}}{\sqrt{s}}-\frac{t}{s})]^{-1} \nonumber\\
\Rightarrow \frac{1}{u}&\approx&-\frac{1}{s} [1+(\frac{ k_{\bot}}{\sqrt{s}}-\frac{t}{s}) \nonumber\\
                       &+&(\frac{ k_{\bot}}{\sqrt{s}}-\frac{t}{s})^2 \nonumber\\
                       &+&(\frac{ k_{\bot}}{\sqrt{s}}-\frac{t}{s})^3 \nonumber\\
                       &+&(\frac{ k_{\bot}}{\sqrt{s}}-\frac{t}{s})^4 \nonumber\\
                       &+&(\frac{ k_{\bot}}{\sqrt{s}}-\frac{t}{s})^5+.......
\label{eq30}
\eea
The binomial expansion of $[1-(\frac{ k_{\bot}}{\sqrt{s}}-\frac{t}{s})]^{-1}$ converges if
$(\frac{k_{\bot}}{\sqrt{s}}-\frac{t}{s})<1$. For the kinematic limit mentioned
above {\it i.e.} for $k_{\bot}\rightarrow0$ and keeping terms
upto  $\mathcal{O}\mathrm{(\frac{t^3}{s^3})}$, the inequality $(\frac{k_{\bot}}{\sqrt{s}}-\frac{t}{s})< 1$ is satisfied.
We have checked  that terms beyond
$(\frac{ k_{\bot}}{\sqrt{s}}-\frac{t}{s})^5$ in the expression of $\frac{1}{u}$ are not required for the kinematic
limit under consideration.
%In this expansion we keep terms upto those quadratic in $k_{\bot}$. This term, on being multiplied with $\frac{1}{k_{\bot}^2}$
%in the outside (referred to equation ~\ref{matelsq}), will generate a term devoid of $k_{\bot}$. No term with higher exponent
%(i.e. $>$2) of $k_{\bot}$ is needed because that will generate a term at least linear in $k_{\bot}$ on being multiplied
% with the $k_{\bot}^2$ outside; and we chose to take $k_{\bot}\rightarrow0$.
With all these we get,
\bea
\frac{1}{u}&=&-\frac{1}{s} [(1-\frac{t}{s}+\frac{t^2}{s^2}-\frac{t^3}{s^3})\nonumber\\
           &+&(\frac{1}{\sqrt{s}}-\frac{2t}{s\sqrt{s}}+\frac{3t^2}{s^2\sqrt{s}})k_{\bot} \nonumber\\
           &+&(\frac{1}{s}-\frac{3t}{s^2}+\frac{6t^2}{s^3})k_{\bot}^2].
\label {onebyu}
\eea
%%%%%%%%%%%%%%%%%%%%%%%%%%%%%%%%%%%%%%%%%%%%%%%%%%%%%%%%%%%%%%%%%%%%%%%%%%%%%%%%%%%%%%%%%%%%%%%%%%%%%%%%%%%%
Similarly $1/u^2$  can be written as
\bea
\frac{1}{u^2}&=&\frac{1}{s^2}[(1-\frac{2t}{s}+\frac{3t^2}{s^2}-\frac{4t^3}{s^3})\nonumber\\
             &+&(\frac{2}{\sqrt{s}}-\frac{6t}{s\sqrt{s}}+\frac{12t^2}{s^2\sqrt{s}})k_{\bot} \nonumber\\
             &+&(\frac{3}{s}-\frac{12t}{s^2}+\frac{30t^2}{s^3})k_{\bot}^2].
\label {onebyu2}
\eea
%%%%%%%%%%%%%%%%%%%%%%%%%%%%%%%%%%%%%%%%%%%%%%%%%%%%%%%%%%%%%%%%%%%%%%%%%%%%%%%%%%%%%%%%%%%%%%%%%%%%%%%%%%%%
For the assumed kinematic conditions $u^4/s^4$ can be expressed as follows:
\bea
\frac{u^4}{s^4}&=&[(1+\frac{4t}{s}+\frac{6t^2}{s^2}+\frac{4t^3}{s^3}) \nonumber\\
             &-&(\frac{4}{\sqrt{s}}+\frac{12t}{s\sqrt{s}}+\frac{12t^2}{s^2\sqrt{s}})k_{\bot} \nonumber\\
             &+&(\frac{6}{s}+\frac{12t}{s^2}+\frac{6t^2}{s^3})k_{\bot}^2].
\label {u4bys4}
\eea
%%%%%%%%%%%%%%%%%%%%%%%%%%%%%%%%%%%%%%%%%%%%%%%%%%%%%%%%%%%%%%%%%%%%%%%%%%%%%%%%%%%%%%%%%%%%%%%%%%%%%%%%%%%%
Similarly,
\bea
\frac{u^3}{s^3}&=&-[(1+\frac{3t}{s}+\frac{3t^2}{s^2}+\frac{t^3}{s^3})\nonumber\\
               &-&(\frac{3}{\sqrt{s}}+\frac{6t}{s\sqrt{s}}+\frac{3t^2}{s^2\sqrt{s}})k_{\bot} \nonumber\\
               &+&(\frac{3}{s}+\frac{3t^2}{s^2})k_{\bot}^2];
\label {u3bys3}
\eea
%%%%%%%%%%%%%%%%%%%%%%%%%%%%%%%%%%%%%%%%%%%%%%%%%%%%%%%%%%%%%%%%%%%%%%%%%%%%%%%%%%%%%%%%%%%%%%%%%%%%%%%%%%%%
and
\bea
\frac{u^2}{s^2}&=&[(1+\frac{2t}{s}+\frac{t^2}{s^2})-(\frac{2}{\sqrt{s}}+\frac{2t}{s\sqrt{s}})k_{\bot} \nonumber\\
               &+&\frac{1}{s}k_{\bot}^2]
\label {u2bys2}.
\eea
Putting Eqs. ~\ref{onebyu} to ~\ref{u2bys2} in
~\ref{matelsq} we get,
\bea
{|M|^2}_{\mathrm{gg\rightarrow ggg}}&=& 12g^2 |{M _{\mathrm{gg\rightarrow gg}}}|^2\frac{1}{k_{\bot}^2} \nonumber\\
&\times&[(1+\frac{t}{2s}+\frac{5t^2}{2s^2}-\frac{t^3}{s^3})\nonumber\\
&-&(\frac{3}{2\sqrt{s}}+\frac{4t}{s\sqrt{s}}-\frac{3t^2}{2s^2\sqrt{s}})k_{\bot} \nonumber\\
&+&(\frac{5}{2s}+\frac{t}{2s^2}+\frac{5t^2}{s^3})k_{\bot}^2)].
\label {matelfinal}
\eea
The terms $\mathcal{O}(k_{\perp}^{-1})$ and 
$\mathcal{O}(k_{\perp}^{0})$  contribute
to the energy loss of the gluons in a gluonic plasma and
hence are important for heavy-ion phenomenology at
RHIC and LHC energies. These terms were absent in the
previous work~\cite{GR} (also in~\cite{DA})
%But within their approximation, considered in the present discussion also,
%two more terms were left, viz. those linear and quadratic with
%$k_{\bot}$. These terms, on being multiplied with $\frac{1}{k_{\bot}^2}$ will give three terms of
%$\mathcal{O}(\frac{1}{\mathrm{k}_{\bot}^2}), \mathcal{O}(\frac{1}{\mathrm{k}_{\bot}^1})$ and
%$\mathcal{O}(\frac{1}{\mathrm{k}_{\bot}^0})$
%--- the most three significant terms within the given approximation.
%*****************************************************************
\section{Acknowledgment:}
JA and SKD are partially supported by DAE-BRNS project
Sanction No.  2005/21/5-BRNS/245.
%%%%%%%%%%%%%%%%%%%%%%%%%%%%%%%%%%%%%%%%%%%%%%%%%%%%%%%%%%%%%%%%%%%%%%%%%%%%%%%%%%%%%%%%%%%%%%%%%%%%%%%%%%%%%%%%%%%%%%%%%%%%%%%%%

\end{document}